\documentclass[12pt, oneside]{article}

\usepackage[]{graphicx}
\usepackage[]{color}
\makeatletter
\def\maxwidth{ %
  \ifdim\Gin@nat@width>\linewidth
    \linewidth
  \else
    \Gin@nat@width
  \fi
}
\makeatother

\definecolor{fgcolor}{rgb}{0.345, 0.345, 0.345}

\usepackage{framed}
\makeatletter
 {\par\unskip\endMakeFramed%
 \at@end@of@kframe}
\makeatother

\definecolor{shadecolor}{rgb}{.97, .97, .97}
\definecolor{messagecolor}{rgb}{0, 0, 0}
\definecolor{warningcolor}{rgb}{1, 0, 1}
\definecolor{errorcolor}{rgb}{1, 0, 0}
\usepackage{alltt}
\usepackage{amsmath}
\usepackage{amsthm}
\usepackage[a4paper]{geometry}
\usepackage{graphicx}
\usepackage{microtype}
\usepackage{siunitx}
\usepackage{booktabs}
\usepackage{cleveref}
\usepackage{float}
\restylefloat{table}

\usepackage{bm}
\usepackage{pgfplots}
\usepackage{pgfplotstable}
\usepackage{array}
\usepackage{colortbl}
\usepackage{rotating}
\usepackage{graphicx,rotating,booktabs}
\usepackage{authblk}

\usepackage[verbose]{placeins}
\usepackage{blindtext}
\usepackage{longtable}
\usepackage{pdflscape}

\usepackage{pdflscape}

\usepackage{graphicx}
\usepackage{endfloat}
\usepackage{amsfonts}
\usepackage{mychicago}
\usepackage{subfigure}
\usepackage{booktabs}

\usepackage{multirow}
\usepackage{caption}
\usepackage{tabu}

\usepackage{cite}

\usepackage[justification=centering]{caption}

\usepackage{setspace}
\usepackage{algorithm2e}

\DeclareMathOperator{\vech}{vech}

\DeclareMathOperator{\tr}{tr}

\makeatletter
      \def\@setcopyright{}
      \def\serieslogo@{}
      \makeatother
\newtheorem*{mydef*}{Definition}

\newtheorem*{mydef2*}{One-sample test statistic}

\newtheorem*{mydef3*}{Two-sample test statistic}

\IfFileExists{upquote.sty}{\usepackage{upquote}}{}
\begin{document}

\title{ Test of Covariance and Correlation Matrices}
\author[1]{\small Longyang Wu }
\author[2]{\small Chengguo Weng}
\author[3]{\small Xu Wang}
\author[4]{\small Kesheng Wang}
\author[5,6]{\small Xuefeng Liu}

\affil[1]{\footnotesize ThermoFisher Scientific, San Clara, CA, 95051}
\affil[2]{\footnotesize Department of Statistics and Actuarial Science\\
	 University of Waterloo,  Waterloo, Canada, N2L 3G1}
\affil[3]{\footnotesize Department of Mathematics\\
	Wilfrid Laurier University,  Waterloo, Canada, N2L 3G5 }
\affil[4]{\footnotesize Department of Biostatistics and Epidemiology, East Tennessee State University, Johnson City, TN, 37614}
\affil[5]{\footnotesize Department of Systems, Populations and Leadership, University of Michigan, Ann Arbor, MI, 48109}
\affil[6]{\footnotesize Department of Biostatistics, University of Michigan, Ann Arbor, MI, 48109}

%\date{}
\maketitle

\begin{abstract}
Based on a generalized cosine measure between two symmetric matrices,  we propose a general framework for one-sample and  two-sample tests of covariance and correlation matrices.  We also develop a set of associated permutation algorithms for some common one-sample tests, such as the tests of sphericity, identity and compound symmetry, and the $K$-sample tests of multivariate equality of covariance or correlation matrices.  The proposed method is very flexible in the sense that it does not assume any underlying distributions and  data generation models.  Moreover,  it allows data to have different marginal distributions in both the one-sample identity and $K$-sample tests. Through real datasets and extensive simulations, we demonstrate that the proposed method performs well in terms of empirical type I error and power in a variety of hypothesis testing situations in which data of different sizes and dimensions are generated using different distributions and generation models.

\end{abstract}


%Confidence Interval of the theta between two random matrices! 

\section{Introduction}

Testing covariance and correlation matrices is an enduring topic in statistics. In addition to its apparent importance in classical multivariate analysis such as multivariate analysis of variance, Fisher's linear discriminant analysis, principle component analysis, and repeated measure analysis \shortcite{Johnson2007}, the hypothesis testing of covariance or correlation matrices has a wide range of applications  in many scientific fields such as psychometric analysis of behavior difference  \shortcite{green1992testing},  co-movement dynamics of credit risk \shortcite{annaert2006intertemporal}, genetic analysis of gene expression \shortcite{singh2002gene}, dynamic modeling of the correlation structure of the stock market \shortcite{djauhari2014dynamics}.

%In practice, log-normal distributions, extreme value distributions etc 

%In multivariate financial times series analysis, it is often of interest to test correlation between the same group of stocks change over time.

%Before financial crisis and after the crisis.

%In longitudinal study of biological process, to test correlation structure change over time. 

%In these cases the usual assumption of two sample study, the independence assumption is not valid. 

%Consider testing the following two null hypothses,

%Test equality of covariance matrices or correlation matrices are often the first step in above mentioned multivariate statistical methods.

Traditionally, the tests of covariance or correlation matrices are discussed separately in the literature. \citeN{bartR} proposed a method to test if a sample correlation matrix is an identity matrix.  \citeN{jennrich1970asymptotic} proposed an asymptotic $\chi^2$ test for the equality of two correlation matrices of multivariate normal distributions. Jennrich's test has been adopted in several statistical softwares and applied in financial studies.  Likelihood ratio based tests for covariance matrices of multivariate normal distributions are well documented in the classical multivariate analysis literature (e.g. \citeN{Muirhead2005}, Chp. 8). Since the classical likelihood ratio based tests are generally not applicable in high-dimensional data due to the  singularity of sample covariance matrices, the recent methodology developments focus on the tests of  high-dimensional covariance and correlation matrices, where the dimension $p$ is much bigger than the sample size $n$.  Among others, \citeN{ledoit2002some} studied the performance of two classical statistics for the one-sample test of sphericity and identity in a high-dimensional setting.  \citeN{chen2010tests} adopted the similar test statistics as in \citeN{ledoit2002some} but used U-statistics to estimate the population covariance matrix. \citeN{li2012two} proposed a two-sample equality test of covariance matrices. \citeN{fujikoshi2011multivariate} provided a summary of some classical and high-dimensional tests of covariance matrices of multivariate normal distributions.  \citeN{yao2015sample} discussed the recent developments on estimation and hypothesis tests of covariance matrices in high dimensions from the random matrix theory viewpoint. \citeN{cai2017global} presented an up-to-date review of methodology developments of the  equality test of high-dimensional covariance matrices with sparse signals. 

There obviously lacks a unified and flexible hypothesis testing framework for practitioners to handle both covariance and correlation matrices without switching between different test statistics or procedures. In this study we propose such a unified and easy-to-use approach for many common tests of covariance and correlation matrices.  We exploit a simple geometric property of symmetric matrices to construct test statistics and develop the associated permutation algorithms to compute p-values under a variety of null hypotheses. Our proposed method can handle a variety of  one-sample tests including the tests of sphericity, identity and compound symmetry, and two- and $K$-sample equality tests ($K>2$). One advantage of our proposed sphericity and compound symmetry tests is that the proposed method does not involve either the unknown  population variance or the intra-class correlation coefficient.  Moreover, the proposed method is very flexible in the sense that it does not assume any underlying distributions or data generation models.  Through both real data analyses and extensive simulations, we demonstrate that the proposed method  can control the type I error at the pre-specified nominal level and provide good power under the alternative in both one-sample and $K$-sample tests, even for high-dimensional data where the dimension $p$ is much bigger than the sample size $n$.

This paper is organized as follows. Section \ref{sec:test-stats} introduces a generalized cosine measure between two symmetric matrices and, based on this notion of generalized cosine, the test statistics for one-sample and two-sample tests  are proposed.  Section \ref{sec:one-sample} studies the one-sample tests  of sphericity, identity  and compound symmetry. Section \ref{sec:k-sample} discusses the two-sample equality test of covariance or correlation matrices and extends the method to handle the $K$-sample problem. Section \ref{sec:simulation} contains extensive simulation studies on the performance of the proposed one-sample and two-sample tests.  In addition, Section \ref{sec:simulation} probes into factors that affect the performance of the proposed methods. The paper concludes with discussions in Section \ref{sec:discussion}.

\section{Test statistic} \label{sec:test-stats}

\label{key}
Below is the definition of the generalized cosine measure between two symmetric matrices. 

\begin{mydef*}\label{def:cos}
Let  $\mathbb{R}^{p\times p}$ be a vector space of symmetric matrices of size $p\times p$ and $f$ :  $\mathbb{R}^{p\times p}$ $\rightarrow$  $\mathbb{R}^{m \times n}$. The angle between two symmetric matrices $\bm{M}_1$ and $\bm{M}_2$ in $R^{p\times p}$ with respect to $f$ is defined as $arccos(\bm{M}_1,\bm{M}_2)$ and
\begin{equation}
cos(\bm{M}_1,\bm{M}_2)=\frac{\langle f(\bm{M}_1), f(\bm{M}_2) \rangle}{\|f(\bm{M}_1)\|\|f(\bm{M}_2)\|},  \label{eqn:def}
\end{equation}
where $\langle\cdot, \cdot\rangle$  is an inner product and $\|\cdot\|$ is the corresponding norm in $\mathbb{R}^{m \times n}$. 
\end{mydef*}

We use two symmetric matrices $\bm{A}$ and $\bm{B}$ to illustrate some possible cosine values with different mappings, their associated inner products and norms.  Matrix $\bm{A}$ is 
\begin{equation} \notag
\begin{bmatrix}
1.00  & 0.50 &  0.33&  0.25 & 0.20 \\
0.50 & 1.00 & 0.25 & 0.20  &0.17 \\
0.33 & 0.25 & 1.00 & 0.17  &0.14 \\
0.25 & 0.20 & 0.17 & 1.00  &0.12 \\
0.20 & 0.17 & 0.14 & 0.12  &1.00
\end{bmatrix},
\end{equation}
which is a modified Hilbert matrix defined as follows:
\begin{equation} \label{eqn:hilbert}
\bm{A}\left(i,j\right)=\begin{cases}
\frac{1}{i+j-1}, & i \neq j,\\
1, & i = j,
\end{cases}
\end{equation} 
for $i, j = 1, \ldots 5$.  Matrix $\bm{B}$  is 
\begin{equation} \notag
\begin{bmatrix}
1.00& 0.74& 0.83& 0.54& 0.41 \\
0.74& 1.00& 0.55& 0.60& 0.34 \\
0.83& 0.55& 1.00& 0.28& 0.58 \\
0.54& 0.60& 0.28& 1.00& 0.48 \\
0.41& 0.34& 0.58& 0.48& 1.00
\end{bmatrix},
\end{equation}
which is a realization of the random matrix  $\bm{A^*}$,  constructed from $\bm{A}$, as follows:
\begin{equation} \label{eqn:b}
\bm{A}^*\left(i,j\right)=\bm{A}^*\left(j,i\right) =\begin{cases}
\frac{1}{i+j-1} + \epsilon, & i \neq j, \\
1, & i =j,
\end{cases}
\end{equation}
where $\epsilon$ is generated from a uniform distribution over the interval (0, 0.5).

Case 1: Suppose $f$ is an identity mapping, i.e., $f(\bm{A})=\bm{A}$.  We consider the Frobenius inner product $\langle \bm{A},\bm{B}  \rangle_F =\tr(\bm{A}^T\bm{B})$ and its corresponding norm $\| \bm{A}\|_F = \tr(\bm{A}^T\bm{A})^{\frac{1}{2}}$. By Equation (\ref{eqn:def}), the cosine between $\bm{A}$ and $\bm{B}$ is calculated as 
\begin{equation}
cos(\bm{A}, \bm{B}) = \frac{\tr(\bm{A}^T\bm{B})}{\tr(\bm{A}^T\bm{A})^{\frac{1}{2}}\tr(\bm{B}^T\bm{B})^{\frac{1}{2}}},
\end{equation}
which is 0.92,  corresponding to an angle of 0.41 radian or $\ang{23}$.  

Case 2:  In this case, we construct a two-step mapping $f$ : $\mathbb{R}^{p\times p}$ $\rightarrow$  $\mathbb{R}^{m}$. In step one we apply the Cholesky decomposition to obtain $\bm{A}=\bm{L}_1\bm{L}_1^T$ and $\bm{B}=\bm{L}_2\bm{L}_2^T$, where $\bm{L}_1$ and $\bm{L}_2$ are lower triangle matrices with positive diagonal elements. In step two we apply the half-vectorization operator (which vectorizes or stacks only the lower trianguler portion of a symmetric matrix) to $\bm{L}_1$ and $\bm{L}_2$ separately to obtain two vectors $\vech(\bm{L}_1)$ and $\vech(\bm{L}_2)$, each with a length of  $m=p(p+1)/2$. To apply Equation (\ref{eqn:def}),  we adopt the Euclidean inner product and norm to compute cosine  as 
\begin{equation}
cos(\bm{A}, \bm{B}) = \frac{\vech(\bm{L_1})^T\vech(\bm{L_2})}{\|\vech(\bm{L}_1)\| \|\vech(\bm{L}_2)\|},
\end{equation}
which is 0.87, corresponding to an angle of 0.52 radian or $\ang{29.5}$. This mapping is feasible only if $\bm{A}$ and $\bm{B}$ are positive definite matrices. 

%Note that for positive definite matrix the Cholesky decomposition is unique.  

Case 3: The mapping $f$ can also be constructed by first applying the eigen-decomposition  to obtain $\bm{Q}_1 \bm{\Lambda}_1\bm{Q}_1^T$ for $\bm{A}$ and $\bm{Q}_2 \bm{\Lambda}_2\bm{Q}_2^T$ for $\bm{B}$, respectively. $\bm{Q}_1$ and $\bm{Q}_2$ are orthogonal matrices of eigenvectors, while $\bm{\Lambda}_1$ and $\bm{\Lambda}_2$ are  diagonal matrices whose elements are the eigenvalues of $\bm{A}$ and $\bm{B}$, respectively. Provided that not all diagonal elements in $\bm{\Lambda}_1$ and $\bm{\Lambda}_2$ are zero,  we can extract the diagonal elements from $\bm{\Lambda_1}$ and $\bm{\Lambda_2}$ to create vectors $v_1$ and $v_2$, respectively.  Thus, we construct a mapping $f$ : $\mathbb{R}^{p\times p}$ $\rightarrow$  $\mathbb{R}^{p}$. To apply Equation (\ref{eqn:def}), we adopt the Euclidean inner product and norm to compute the cosine as 
\begin{equation}
cos(\bm{A}, \bm{B}) = \frac{v_1^T v_2}{\|v_1\| \|v_2\|},
\end{equation}
which is 0.93, corresponding to a angle of 0.37 radian or approximately $\ang{21}$.

Case 4:  Since a $p\times p$ symmetric  matrix is completely determined by its lower triangular elements together with the symmetry, we construct the mapping $f$ by applying the half-vectorization operator on $\bm{A}$ and $\bm{B}$ directly.  The cosine between $\bm{A}$ and $\bm{B}$ can be obtained from Equation (\ref{eqn:def}) as follows:
\begin{equation}
cos(\bm{A}, \bm{B}) = \frac{\vech(\bm{A})^T\vech(\bm{A})}{\|\vech(\bm{A})\| \|\vech(\bm{B})\|},
\end{equation}
which is 0.94,  corresponding to a angle of 0.35 radian or $\ang{20}$.  

For correlation matrices, the computation can be simplified by introducing a modified half-vectorization operator $\vech^*(\cdot)$ from $\vech(\cdot)$ by excluding the diagonal elements of the matrix.  Suppose  $\bm{A}$ and $\bm{B}$ are two correlation matrices, the cosine  
can be computed by Equation (\ref{eqn:def}) as 
\begin{equation}
cos(\bm{A}, \bm{B}) = \frac{\vech^*(\bm{A})^T\vech^*(\bm{A})}{\|\vech^*(\bm{A})\| \|\vech^*(\bm{B})\|},
\end{equation}
which is 0.95, corresponding to 0.31 radian or $\ang{18}$. 

In summary,  we have considered four different mappings to compute the generalized cosine between  the two modified Hilbert matrices $\bm{A}$ and $\bm{B}$. The mapping based on the Cholesky decomposition gives the lowest cosine value of 0.87.  The mappings based on the Frobenius inner product and the eigen-decomposition give the similar cosine values of 0.92 and 0.93, respectively. When the redundant information is removed, the half-vectorization and the modified half-vectorization operators give the largest cosine values of 0.94 and 0.95, respectively.  In this study we choose the half-vectorization operator, $\vech(\cdot)$, for covariance matrices and the modified half-vectorization operator, $\vech^*(\cdot)$, for correlation matrices,  because these two operators completely remove the redundancy in symmetric matrices and are very easy to compute. Moreover, these two operators demonstrate better statistical  power in our pilot simulation studies. The cosine value computed from Equation (\ref{eqn:def})  measures the similarity between two symmetric matrices. When this value is one, the two matrices are identical. We exploit this property to propose new test statistics for one-sample and two-sample tests of covariance and correlation matrices.
 
\begin{mydef2*}\label{def:test1}
 	Let $\bm{S}$ be the sample covariance (correlation) matrix with the corresponding population covariance (correlation) matrix $\bm{\Sigma}$.   The test statistic for the equality of  $\bm{\Sigma}$ and $\bm{\Sigma}_0$, which is either known or specified up to some unknown parameters, is 	
 	\begin{equation} \label{eqn:one_sample}
 	1-cos(\bm{S}, \bm{\Sigma}_0),
 	\end{equation}
 	where the cosine is computed according to Equation (\ref{eqn:def}) under a pre-determined mapping $f$ with a properly chosen  inner product and norm.
\end{mydef2*}

\begin{mydef3*}\label{def:test2}
Let $\bm{S}_1$  and $\bm{S}_2$ be the sample covariance (correlation) matrices with the corresponding population covariance (correlation) matrices $\bm{\Sigma}_1$ and $\bm{\Sigma}_2$, respectively.   The test statistic for the equality of  $\bm{\Sigma}_1$ and $\bm{\Sigma}_2$ is 	
\begin{equation} \label{eqn:two_sample}
1-cos(\bm{S}_1, \bm{S}_2),
\end{equation}
where the cosine is computed according to Equation (\ref{eqn:def}) under a pre-determined mapping $f$ with a properly chosen inner product and norm.
\end{mydef3*}

It is worth pointing out that the two newly proposed test statistics exploit a simple geometric property of symmetric matrices and require no distributional assumptions of data. We further adopt a permutation approach to obtain the null distributions of the two test statistics and compute p-values under the null hypotheses of different tests.
% Moreover, 

% and underlying data generation model.  

%and can handle different types of correlation matrices, for example, the %Spearman or Kendall rank correlation matrices.    

\section{One-sample test} \label{sec:one-sample}

Suppose a matrix, $\bm{D}_{n\times p}=\{ x_{ij}\}$, $i=1,\ldots, n$ and $j=1,\ldots, p$, contains data collected on $n$ subjects and each with $p$ covariates.  Its $i$th row is a realization of random vector  $\bm{X} = (X_1, X_2, \ldots, X_p)^T$, which has an unknown distribution with a covariance matrix $\bm{\Sigma}$ and a correlation matrix $\bm{R}$. In the one-sample problem one may use the data matrix  $\bm{D}_{n\times p}$ to test if  the covariance matrix $\bm{\Sigma}$ is proportional to an identity matrix,  known as the sphericity test,  or  to test if $\bm{\Sigma}$ or $\bm{R}$ equals to an identity matrix,  known as the identity test.

\subsection{Sphericity test} \label{sec:sph}
The hypotheses in the sphericity test  for a covariance matrix are as follows: 
\begin{equation} \label{eqn:shp}
H_0: \bm{\Sigma} = \sigma^2\bm{I}   \;\;  \mathrm{vs.}  \;\; H_1: \bm{\Sigma} \neq \sigma^2\bm{I},
\end{equation}
where $\sigma^2$ is the unknown population variance under the null. In this case, $\bm{\Sigma}_0$ in the one-sample test statistic (\ref{eqn:one_sample}) is $\sigma^2\bm{I}$ and $\bm{S}$ is the sample covariance  matrix of the data matrix $\bm{D}_{n \times p}$.  To test the sphericity of a covariance matrix,  using $\vech(\cdot)$, the Euclidean inner product and norm,  the test statistic (\ref{eqn:one_sample}) can be shown to be
\begin{equation} \label{eqn:sph_var}
1-\frac{\tr(\bm{S})}{\sqrt{p}\|\vech(\bm{S})\|},
%1-\frac{\vech(\bm{S})^T\vech(\bm{I})}{\sqrt{p}\|\vech(\bm{S})\|}.
%1-\frac{\sum_{i=1}^{p}s_{ii}^2}{\sqrt{p}\|\vech(\bm{S})\|},
\end{equation} 
where $\tr(\cdot)$ denotes the trace. The derivation of (\ref{eqn:sph_var}) is provided in Appendix I. Note that the test statistic (\ref{eqn:sph_var}) does not depend on the unknown parameter $\sigma^2$.  

The sphericity test (\ref{eqn:shp}) is a combination of two tests \shortcite{fujikoshi2011multivariate}: 1)  $H_0: \bm{\Sigma} = \sigma^2\bm{I}$ vs. $H_1: \bm{\Sigma} = \mathrm{diag}(\sigma_1^2, \sigma_2^2, \ldots, \sigma_p^2)$ and 2) $H_0:  \bm{\Sigma} = \mathrm{diag}(\sigma_1^2, \sigma_2^2, \ldots, \sigma_p^2)$ vs. $H_1: \bm{\Sigma} \neq  \mathrm{diag}(\sigma_1^2, \sigma_2^2, \ldots, \sigma_p^2)$.   To handle these two tests simultaneously,  we propose Algorithm \ref{alg:sph} to compute p-values under the null of sphericity.  This algorithm first permutes the elements in each row of $\bm{D}_{n \times p}$ to generate a permuted data matrix $\bm{D}^{\dagger}_{n \times p}$, then the elements in each column of $\bm{D}^{\dagger}$ to generate $\bm{D}^{\ddagger}_{n \times p}$.  The permutation on each row of the data matrix is to test if the diagonal elements in $\bm{\Sigma}$ are identical; the permutation on each column is to test if the off-diagonal elements of $\bm{\Sigma}$ are zero.  Algorithm 1 assumes that the elements in the random vector $\bm{X}$ are exchangeable in the sense that $\bm{\Sigma}$ is invariant to the permutations of the elements in $\bm{X}$ under the null hypothesis of (\ref{eqn:shp}). This algorithm requires that the elements in $\bm{X}$ have the same distribution so that the elements in each row of  $\bm{D}_{n \times p}$ can be permuted.
 
 \begin{center}
	\RestyleAlgo{boxruled}
	\LinesNumbered	
	\begin{algorithm}[ht]
		\caption{Test the sphericity of a covariance  matrix.} \label{alg:sph}
		%\vspace*{-12pt}
		\begin{tabbing}
			\enspace  $r$  $\leftarrow$  the number of permutations \\
			%\enspace Set $T(i)=0$, $i=1,\ldots, r$ \\
			\enspace  $T(i)  \leftarrow 0$, $i=1,\ldots, r$ \\
			\enspace $\bm{S} \leftarrow $ Compute the sample covariance matrix of $\bm{D}_{n \times p}$ \\
			\enspace  $T_o \leftarrow $  Compute the test statistic (\ref{eqn:sph_var})  using $\bm{S}$\\
			\enspace For $i=1$ to $i=r$ \\
			\qquad $\bm{D}^{\dagger}_{n \times p} \leftarrow$  Permute the elements in each row of $\bm{D}_{n \times p}$ \\
			\qquad $\bm{D}^{\ddagger}_{n \times p} \leftarrow$  Permute the elements in each column of $\bm{D}^{\dagger}_{n \times p}$ \\
			\qquad $\bm{S}^* \leftarrow $ Compute the sample covariance matrix of $\bm{D}^{\ddagger}_{n \times p}$ \\
			\qquad  $T(i) \leftarrow $  Compute the test statistic (\ref{eqn:sph_var})  using $\bm{S}^*$\\ 				
			\enspace End For\\
			\enspace Report $p$-value $ \leftarrow (\#(T(i) \ge T_o)+1)/(r+1)$
		\end{tabbing}
	\end{algorithm}
\end{center}

We illustrate the proposed sphericity test and Algorithm \ref{alg:sph} using the \verb|bfi| data in the R package \verb|pysch| \shortcite{psych}. The  \verb|bfi| data contain twenty-five personality self reported items from 2800 subjects who participated in the Synthetic Aperture Personality Assessment (SAPA) web based personality assessment project. These twenty-five items are organized by five putative factors: Agreeableness, Conscientiousness, Extraversion, Neuroticism, and Openness. Each subject answered these personality questions using a six point response scale ranging from 1 (Very Inaccurate) to 6 (Very Accurate). After removing the missing data,  we apply the proposed sphericity test and Algorithm \ref{alg:sph} to test the sphericity of  the $25\times25$ covariance matrix of  personality measurements of 2436 subjects. We obtain a p-value of  0.01 based on 100 permutations using Algorithm \ref{alg:sph}. Therefore, the null hypothesis of sphericity is rejected at the significance level of 0.05.  We will further investigate the empirical type I error and power of our proposed sphericity test in Section \ref{sec:simulation}. 

\subsection{Identity test}

The identity test of a covariance matrix can be formulated as \cite{ledoit2002some}
\begin{equation} \label{eqn:var_idt_H}
H_0 : \bm{\Sigma} = \bm{I} \; \mathrm{vs.} \; H_1 : \bm{\Sigma} \neq  \bm{I},
\end{equation}
which includes the more general null hypothesis $H_0: \bm{\Sigma}= \bm{M}_0$, where $\bm{M}_0$ is a known covariance matrix,  since we can replace $\bm{M}_0$ by $\bm{I}$  and  multiply data by $\bm{M}_0^{-1/2}$.   

Because $\bm{I}$ in the hypothesis (\ref{eqn:var_idt_H}) and $\sigma^2\bm{I}$ in the sphericity hypothesis (\ref{eqn:shp}) only differ by a constant $\sigma^2$ and the test statistic (\ref{eqn:sph_var}) does not depend on $\sigma^2$, we can adopt the test statistic (\ref{eqn:sph_var}) to test the hypothesis (\ref{eqn:var_idt_H}). %Furthermore, unlike the sphericity test,  the identity test only focuses on off-diagonal elements of the sample covariance matrix $\bm{S}$ of $\bm{D}_{n \times p}$ and the test statistic (\ref{eqn:sph_var}) can be shown to be
%\begin{equation} \label{eqn:idt_var}
%1-\frac{\vech(\bm{S})^T\vech(\bm{I})}{\sqrt{p}\|\vech(\bm{S})\|}.
%1-\frac{\sum_{i=1}^{p}s_{ii}^2}{\sqrt{p}\|\vech(\bm{S})\|}.
%\end{equation} 

For the identity test of a correlation matrix, 
\begin{equation} \label{eqn:r_idt_H}
H_0 : \bm{R} = \bm{I} \;  \mathrm{vs.}  \; H_1 : \bm{R} \neq \bm{I},
\end{equation}
for which, the  test statistic (\ref{eqn:sph_var}) can be simplified into
\begin{equation} \label{eqn:idt_R}
1-\frac{\sqrt{p}}{\|\vech(\bm{S})\|}.
\end{equation} 
The derivation of   (\ref{eqn:idt_R}) is provided in Appendix I.

Since the identity test mainly concerns the off-diagonal elements of a covariance or correlation matrix, we propose Algorithm \ref{alg:idt}, which permutes the elements in each column of the data matrix $\bm{D}_{n \times p}$, to compute p-values under the null. Permuting the elements in columns can disrupt the covariance structure, but it does not change the sample variance of each column of the data matrix, i.e. the diagonal elements of the sample covariance matrix remain unchanged in different permutations. This implies that, for a given $\bm{D}_{n \times p}$, $\tr(\bm{S})$ remains a constant in (\ref{eqn:sph_var}) in different permutations. For a correlation matrix $\bm{S}$, we have $\tr(\bm{S})$=$p$. Since the column-wise  permutations can sufficiently disassociate every pair of random variables, and hence, disrupt the covariance structure of the $p$ random variables, the off-diagonal elements of the sample covariance matrix of the permuted data are much closer to zero than the original sample covariance (correlation) matrix  under the alternative. Therefore, the sample covariance and correlation matrices with a good portion of non-zero off-diagonal elements have larger values of the statistics of (\ref{eqn:sph_var}) and (\ref{eqn:idt_R}), respectively, to reject the null hypothesis of identity.  Moreover, Algorithm \ref{alg:idt} does not require each column of a data matrix to have the same distribution.

\begin{center}
	\RestyleAlgo{boxruled}
	\LinesNumbered	
	\begin{algorithm}[ht]
		\caption{Test the identity of  covariance or correlation matrix.} \label{alg:idt}
		%\vspace*{-12pt}
		\begin{tabbing}
			\enspace $r$  $\leftarrow$  the number of permutations  \\
			\enspace $T(i)$ $\leftarrow 0$, $i=1,\ldots, r$ \\
			\enspace $\bm{S} \leftarrow $    Compute the sample covariance or correlation matrix of $\bm{D}_{n \times p}$ \\
			\enspace  $T_o \leftarrow $ Compute the test statistic (\ref{eqn:sph_var}) or (\ref{eqn:idt_R}) using $\bm{S}$ \\
			\enspace For $i=1$ to $i=r$ \\
			\qquad $\bm{D}^{*}_{n \times p} \leftarrow$ Permute the elements in each column of $\bm{D}_{n \times p}$ \\
			\qquad $\bm{S}^{*} \leftarrow $ Compute the sample covariance or correlation matrix of $\bm{D}^{*}_{n \times p}$ \\
			\qquad  $T(i) \leftarrow $ Compute  the test statistic (\ref{eqn:sph_var}) or (\ref{eqn:idt_R})  using $\bm{S}^*$   \\
			
			\enspace End For\\
			\enspace Report $p$-value $ \leftarrow (\#(T(i) \ge T_o)+1)/(r+1)$
		\end{tabbing}
	\end{algorithm}
\end{center}

%If $\bm{R}$ is a Pearson's correlation matrix, then the null hypotheses of (\ref{eqn:var_idt_H}) and (\ref{eqn:r_idt_H}) are equivalent in the sense that acceptance or rejection of  (\ref{eqn:var_idt_H}) will result in the same action for (\ref{eqn:r_idt_H}).   
In addition to the Pearson's correlation matrix, the test statistic defined in (\ref{eqn:idt_R}) and its associated Algorithm \ref{alg:idt} can handle the  Spearman's and  the Kendall's rank correlation matrices, because this test statistic is based on the cosine measure between a sample correlation matrix $\bm{S}$ and $\bm{I}$, and does not require any particular types of  correlation matrices.

We use the \verb|bfi|  data as described in section \ref{sec:sph} to illustrate the proposed identity test and its associated Algorithm \ref{alg:idt} for correlation matrices. For a Pearson's correlation matrix of twenty-five personality self reported items,  Algorithm \ref{alg:idt} produces a p-value of 0.01 based on 100 permutations.  The Bartlett's identity test implemented in the same R package gives a p-value of zero. We also apply the identity test to the Spearman's rank and Kendell's rank correlation matrices.  Algorithm \ref{alg:idt} produces a  p-value of 0.01 for both types of correlation matrices based on 100 permutations .  Therefore, we reject the identity of these three types of correlation matrices. We will further study the performance of our proposed identity test in Section \ref{sec:simulation}.

\subsection{Compound symmetry test}

The hypothesis of compound symmetry is $H_0: \bm{\Sigma}=\sigma^2[(1-\rho)\bm{I} +\rho \bm{J}]$ for a covariance matrix or $H_0: \bm{R}=[(1-\rho)\bm{I} +\rho\bm{J}]$ for a correlation matrix, where $\bm{J}$ is a square matrix of ones and $\rho > 0$ is the intra-class correlation coefficient. In this case,  $\Sigma_0$ in the one-sample test statistic (\ref{eqn:one_sample}) is $\sigma^2[(1-\rho)\bm{I} +\rho \bm{J}]$ for a covariance and $[(1-\rho)\bm{I} +\rho\bm{J}]$ for a correlation matrix, respectively. Using the modified half-vectorization operator, $\vech^*(\cdot)$, a unified test statistic for testing the compound symmetry of both covariance and correlation matrices can be obtained from (\ref{eqn:one_sample}) as

%We propose a test statistic which does not depend on the unknown parameters  $\sigma^2$ and $\rho$ as follows: 

\begin{equation} \label{eqn:com}
1-\frac{\vech^*(\bm{S})^T\vech^*(\bm{J})}{\sqrt{0.5p(p-1)}\|\vech^*(\bm{S})\|},
\end{equation} 
where $\bm{S}$ is either a sample covariance or correlation matrix.  The derivation of (\ref{eqn:com}) is provided in Appendix II. Note that this test statistic does not depend on the unknown parameters $\sigma^2$ and $\rho$.

Suppose a data matrix $\bm{D}_{n \times p}$ contains rows that are the realizations of a $p \times 1 $ random vector $\bm{X}=(X_1, \ldots, X_p)^T$, in which all $X_i$'s have the same distribution. Under the null hypothesis of compound symmetry, it is true that $Var(X_i) = \sigma^2$, for all $i$'s and $Cov(X_i, X_j) = \sigma^2(1-\rho) $ for all pairs of $i,j$, provided $i \neq j$. This implies that the random permutations of the elements in $\bm{X}$ do not alter its covariance matrix under the assumption of compound symmetry.  Using this fact, we propose Algorithm \ref{alg:com} to compute p-values for the test statistic (\ref{eqn:com}) by randomly permuting the elements in each row of the data matrix. For this algorithm to be applicable, each column of the data matrix must have the same distribution.

 %\textsuperscript{*}\footnotesize{Standard deviation of SSNR}

\begin{center}
	\RestyleAlgo{boxruled}
	\LinesNumbered	
	\begin{algorithm}[ht]
		\caption{Test the compound symmetry of covariance (correlation)  matrix.} \label{alg:com}
			\begin{tabbing}
			\enspace  $r \leftarrow$ the number of permutations  \\
			\enspace  $T(i) \leftarrow 0$, $i=1,\ldots, r$ \\
			\enspace $\bm{S} \leftarrow $ Compute the sample covariance (correlation) matrix of $\bm{D}_{n \times p}$ \\
			\enspace  $T_o \leftarrow $ Compute the test statistic (\ref{eqn:com})  using $\bm{S}$\\
			\enspace For $i=1$ to $i=r$ \\
				\qquad $\bm{D}^*_{n \times p} \leftarrow$  Permute the elements in each row of $\bm{D}_{n \times p}$ \\
			\qquad $\bm{S}^* \leftarrow $ Compute the sample covariance (correlation) matrix of $\bm{D}^*_{n \times p}$ \\
			\qquad $T(i)  \leftarrow $ Compute  the test statistic (\ref{eqn:com})  using $\bm{S}^*$\\
			
			\enspace End For\\
			\enspace Report $p$-value $ \leftarrow (\#(T(i) \ge T_o)+1)/(r+1)$
			\end{tabbing}
		\end{algorithm}
%\vspace*{-30pt}	
\begin{flushleft}
%	\textsuperscript{*}\footnotesize{ All $\bm{S}$'s  off-diagonal element are positive and }
\end{flushleft}	
\end{center}
For an illustrative purpose, we consider the Cork data (\citeN{rencher2012methods}, Table 6.21). Four cork borings, one in each of the four cardinal directions, were taken from each of twenty-eight trees and their weights were measured in integers.  The goal is to test the compound symmetry of the covariance matrix of these weight measurements taken in the four cardinal directions.  Using the modified half-vectorization operator $\vech^*(\cdot)$ and the Euclidean inner product and norm,  we obtain a cosine of 0.99 (0.998) between the sample covariance (correlation) matrix and the compound symmetry, indicating a  strong similarity. We apply our proposed compound symmetry test (\ref{eqn:com}) to this dataset and, using Algorithm \ref{alg:com},  obtain  a p-value of 0.099 for the compound symmetry test of the sample covariance matrix and 0.069 for the sample correlation matrix based on 100 permutations. Hence,  the null hypothesis of compound symmetry can not be rejected at the significance level of 0.05. \citeN{rencher2012methods}, however,  rejected the compound symmetry  based on a $\chi^2$-approximation to a likelihood ratio test using a multivariate normal distribution (\citeN{rencher2012methods}, Example 7.2.3). This discrepancy is due to their inappropriate adoption of the multivariate normal distribution, since three of the four marginal distributions failed to pass the normality test. The test statistic (\ref{eqn:com}) and Algorithm \ref{alg:com}, on the other hand, require no distributional assumptions and are robust to the underlying distribution of Cork borings weights.

%\begin{equation}\label{h0:var}
%\mathrm{H}_0: \,\,\, \bm{\Sigma}=\mathrm{diag}\{\sigma^2_1,\ldots,\sigma^2_p\} 
%\end{equation}

%\begin{equation}\label{h0:r}
%\mathrm{H}_0: \,\,\, \bm{R}=\bm{I} 
%\end{equation}
%in that the hypothesis (\ref{h0:var}) is accepted of rejected if and only if (\ref{h0:r}) is accepted or rejected. Therefore, we focus on the null hypothesis (\ref{h0:r}).

\section{$K$-sample test ($K>2$)} \label{sec:k-sample}

In this section we will first discuss the test statistics and their associated algorithms for the two-sample problem ($K$=2), then extend the method to handle the $K$-sample problem.  Let $\bm{D}_{n_1 \times p}$ and $\bm{D}_{n_2 \times p }$ be two data matrices with sample covariance (correlation) matrices $\bm{S}_1$ and $\bm{S}_2$, respectively. Suppose the rows  of  $\bm{D}_{n_1 \times p}$ are the realizations of a $p \times 1$ random vector $\bm{X}_{1}=(X_{11}, \ldots, X_{1p})^T$ and the rows of $\bm{D}_{n_2 \times p}$ are of random vector $\bm{X}_{2}=(X_{21}, \ldots, X_{2p})^T$. Let $\bm{\Sigma}_1$ ($\bm{R}_1$) be the covariance (correlation) matrix of $\bm{X}_1$ and $\bm{\Sigma}_2$ ($\bm{R}_2$) be the covariance (correlation) matrix of $\bm{X}_2$.

In the two-sample problem,  the null hypotheses are  $H_0: \bm{\Sigma}_1 = \bm{\Sigma}_2$ for covariance matrix and   $H_0: \bm{R}_1 = \bm{R}_2$ for correlation matrix.  We adopt  $\vech(\cdot)$, the Euclidean inner product and norm  in the two-sample test statistic  (\ref{eqn:two_sample}) to test the equality of covariance matrices, leading to a test statistic
\begin{equation}  
1-\frac{\vech(\bm{S}_1)^T\vech(\bm{S}_2)}{\|\vech(\bm{S}_1)\|\|\vech(\bm{S}_2)\|},\label{eqn:twov}
\end{equation}  
and $\vech^*(\cdot)$, the Euclidean inner product and norm to test the equality of  correlation matrices, yielding a test statistic
\begin{equation} 
1-\frac{\vech^*(\bm{S}_1)^T\vech^*(\bm{S}_2)}{\|\vech^*(\bm{S}_1)\|\|\vech^*(\bm{S}_2)\|},\label{eqn:twor}
\end{equation} 
where $\bm{S}_1$ and $\bm{S}_2$ are the sample covariance or correlation matrices.

%Note that the $X_i$'s in $\bm{X}$ and the $Y_i$'s in $\bm{Y}$ need not to have the same marginal distribution.

We propose Algorithm \ref{alg:two} to compute p-values for these test statistics under the null hypothesis of equality.  In this algorithm two data matrices  $\bm{D}_{n_1 \times p}$ and $\bm{D}_{n_2 \times p }$ are stacked to form a new data matrix $\bm{D}_{n \times p}$ ($n=n_1 + n_2$). In each permutation the rows of $\bm{D}_{n \times p}$ are randomly permuted to generate a permuted data matrix $\bm{D}^*_{n \times p}$, which is then split into two data matrices of   $\bm{D}_{n_1 \times p}^*$ and $\bm{D}_{n_2 \times p }^*$ to compute the test statistic (\ref{eqn:twov}) or (\ref{eqn:twor}).  Algorithm \ref{alg:two} assumes that $X_{1i}$ in $\bm{X}_1$ and $X_{2i}$ in $\bm{X}_2$ have the same distribution for all $i$'s. One advantage of our proposed two-sample test is that $X_{1i}$ and $X_{1j}$, $i\neq j$, need not to have the same distribution.

Under the null hypothesis of equality $\bm{\Sigma}_1=\bm{\Sigma}_2=\bm{\Sigma}$, the sample covariance matrices $\bm{S}_1^*$  of $\bm{D}_{n_1 \times p}^*$ and $\bm{S}_2^*$ of  $\bm{D}_{n_2 \times p}^*$  can be considered as the realizations of matrix $\bm{\Sigma}$.  The rationale behind Algorithm \ref{alg:two} is that the cosine value between $\bm{S}_1^*$ and  $\bm{S}_2^*$ is similar to that of  $\bm{S}_1$ and  $\bm{S}_2$ under the null hypothesis and the permutations provide a good control of the type I error. Under the alternative, the repeated random-mixing rows of $\bm{D}_{n_1 \times p}$ and $\bm{D}_{n_2 \times p}$ produce $\bm{S}_1^*$  and $\bm{S}_2^*$ such  that  the cosine value between the two is bigger than that of $\bm{S}_1$ and  $\bm{S}_2$, therefore the test  statistics (\ref{eqn:twov}) and (\ref{eqn:twor}) have good power to reject the null at  a pre-determined significance level.

\begin{center}
\RestyleAlgo{boxruled}
\LinesNumbered	
\begin{algorithm}[!h]
\caption{Test for the equality of two covariance (correlation)  matrices.} \label{alg:two}
%\vspace*{-12pt}
\begin{tabbing}

$r \leftarrow $ the number of permutations \\
    
$T(i)\leftarrow 0$, $i=1,\ldots, r$ \\
   $\bm{S}_1 \leftarrow$ Compute the  sample covariance (correlation ) matrix from $\bm{D}_{n_1 \times  p}$ \\
   $\bm{S}_2 \leftarrow$ Compute the sample covariance (correlation ) matrix from $\bm{D}_{n_2 \times p}$ \\
   
$T_o \leftarrow $  Compute (\ref{eqn:twov}) for covariance or (\ref{eqn:twor})  for  correlation matrix\\
$\bm{D}_{(n_1 + n_2) \times p} \leftarrow $ stack 
$\bm{D}_{n_1 \times p}$ and $\bm{D}_{n_2 \times p}$ \\

For $i=1$ to $i=r$ \\
          \qquad $\bm{D}^*_{(n_1 + n_2) \times p} \leftarrow $  randomly shuffle the rows of $\bm{D}_{(n_1 + n_2) \times p}$  \\
          \qquad $\bm{D}^*_{n_1  \times p}  \leftarrow $ the first $n_1$ rows of $\bm{D}^*_{(n_1 + n_2) \times p}$ \\
          \qquad $\bm{D}^*_{n_2 \times p} \leftarrow $ the remaining $n_2$ rows of $\bm{D}^*_{(n_1 + n_2) \times p}$ \\
        
         \qquad $\bm{S}_1^* \leftarrow$ Compute the sample covariance (correlation ) matrix from $\bm{D}^*_{n_1 \times p }$ \\
         \qquad $\bm{S}_2^* \leftarrow$ Compute the sample covariance (correlation ) matrix from $\bm{D}^*_{n_2 \times p}$ \\
         \qquad $T(i)  \leftarrow $ Compute (\ref{eqn:twov}) for covariance or (\ref{eqn:twor})  for correlation matrix  \\

End For \\
Report $p$-value = $(\#(T(i) \ge T_o)+1)/(r+1)$
\end{tabbing}
\end{algorithm}
\end{center}

We use the flea beetles data  (\citeN{rencher2012methods}, Table 5.5)  to illustrate our proposed two-sample equality test. This dataset contains four measurements of two species of flea beetles, {\it Haltica oleraces} and {\it Haltica cardorum}. To test the equality of two covariance matrices of the four measurements of the two flea beetle species, we apply the proposed two-sample test  (\ref{eqn:twov})  and obtain a p-value of 0.37 using  Algorithm \ref{alg:two} with 100 permutations. Assuming normality and using  $\chi^2$- and  $F$- approximation, the $Box's$ $M$-$test$ gives a similar scale p-value of 0.56  \shortcite{rencher2012methods}. In this case our proposed equality test (\ref{eqn:twov})  agrees with the likelihood ratio based $Box's$ $M$-$test$, and  the null hypothesis of the equality of two covariance matrices is not rejected at the significance level of 0.05.

To test the multivariate equality of several covariance or correlation matrices, we consider 
$H_0: \bm{\Sigma}_1 = \bm{\Sigma}_2 = \cdots =\bm{\Sigma}_K$ for the equality of multiple covariance matrices and $H_0: \bm{R}_1 = \bm{R}_2 = \cdots = \bm{R}_K$      
for the equality of  multiple correlation matrices. We propose the test statistic
\begin{equation} \label{eqn:ksample}
T=\max\{T_{12},T_{13}, \ldots, T_{(K-1)K}\},
\end{equation}
where $T_{ij}$ is the test statistic (\ref{eqn:twov}) or (\ref{eqn:twor}) for the pairwise two-sample comparison of $\bm{\Sigma}_i$ ($\bm{R}_i$) and $\bm{\Sigma}_j$ ($\bm{R}_j$) for all possible unique pairs of $K$ populations. Let $\bm{D}_{n_1\times p}, \bm{D}_{n_2 \times p}, \ldots, \bm{D}_{n_K \times p}$ be the data matrices  from $K$ populations.  Algorithm \ref{alg:ksample} provides a permutation approach to compute p-values under the null hypothesis of the multivariate equality of covariance or correlation matrices.

\begin{center}
	\RestyleAlgo{boxruled}
	\LinesNumbered	
	\begin{algorithm}[!h]
		\caption{Test for the equality of $K$ covariance (correlation)  matrices.} \label{alg:ksample}
		%\vspace*{-12pt}
		\begin{tabbing}
			
			 $r \leftarrow$ the number of permutations \\
			
			 $T(i) \leftarrow 0$, $i=1,\ldots, r$ \\
			
			$T_o \leftarrow $ Compute  the test statistic (\ref{eqn:ksample}) for $\bm{D}_{n_1 \times p}$,  $\bm{D}_{n_2 \times p}$ , $\ldots$, $\bm{D}_{n_K \times p}$\\
			$\bm{D} \leftarrow $ stack $\bm{D}_{n_1 \times p}$,  $\bm{D}_{n_2 \times p}$, $\ldots$, $\bm{D}_{n_K \times p}$\\  
					
			For $i=1$ to $i=r$ \\
			\qquad $\bm{D}^* \leftarrow $  randomly shuffle the rows of $\bm{D}$ \\
			\qquad $\bm{D}^*_{n_1 \times p} \leftarrow $ the first $n_1$ rows of $\bm{D}^*$ \\
			\qquad $\bm{D}^*_{n_2 \times p} \leftarrow $ the second $n_2$ rows of $\bm{D}^*$ \\
		    \qquad \qquad $\vdots$   \qquad \qquad $\vdots$  \qquad \qquad $\vdots$ \\
	    	\qquad $\bm{D}^*_{n_K \times p} \leftarrow $ the remaining $n_K$ rows of $\bm{D}^*$ \\
			\qquad $T(i)  \leftarrow $ Compute the test statistic (\ref{eqn:ksample}) for $\bm{D}^*_{n_1 \times p}$,  $\bm{D}^*_{n_2 \times p}$ , $\ldots$, $\bm{D}^*_{n_K \times p}$\\ 
			End For  \\
			Report $p$-value = $(\#(T(i) \ge T_o)+1)/(r+1)$
		\end{tabbing}
	\end{algorithm}
\end{center}

We use the Rootstock data  (\citeN{rencher2012methods}, Table 6.2) to illustrate  our proposed $K$-sample test. Eight apple trees from each of the six rootstocks  were measured on  four variables: 1) trunk girth at 4 years; 2) extension growth at 4 years; 3) trunk girth at 15 years; 4) weight of the tree above ground at 15 years. The null hypothesis is
\begin{equation}
H_0: \bm{\Sigma}_1 = \bm{\Sigma}_2 = \bm{\Sigma}_3 = \bm{\Sigma}_4 = \bm{\Sigma}_5 = \bm{\Sigma}_6,
\end{equation}
for the equality of six covariance matrices of the six rootstocks. The likelihood ratio based $Box's$ $M$-$test$ yields a p-value of 0.71 and 0.74 based on $\chi^2$- and $F$- approximation \shortcite{rencher2012methods}, respectively.  We apply the multivariate equality test (\ref{eqn:ksample}) to this dataset and obtain a p-value of 0.99 using Algorithm \ref{alg:ksample} with 100 permutations.  For this dataset,  our method agrees with the  $Box's$ $M$-$test$, and the null hypothesis of the multivariate equality of these six covariance matrices is not rejected at the significance level of 0.05.

\section{Simulation studies} \label{sec:simulation}
In this section we carry out extensive simulations to investigate the empirical type I error  and power of our proposed one-sample and two-sample tests. In the one-sample setting we investigate the proposed sphericity test and identity test. In the two-sample setting, we investigate the proposed equality test of two covariance matrices.  We design and conduct these simulation studies with the backdrop of high dimensions. We also probe into factors that may affect the performance of the proposed methods. All simulations are conducted in the R computing environment \shortcite{rsys}.

\subsection{One-sample tests}
\subsubsection{Sphericity test}
In this section we evaluate the empirical type I error and power of  our proposed sphericity test.  We adapt the simulation design and model of \shortciteN{chen2010tests}  for a $p$-dimensional random vector $\bm{X} = \left(X_1, \ldots, X_p\right)^T$ as 
\begin{equation} \label{eqn:chenM}
\bm{X} = \bm{\Gamma} \bm{Z},
\end{equation}
where $\bm{\Gamma}$ is a $p \times m$ constant matrix with $p \leq m$ and  $\bm{Z}=(Z_1, \cdots, Z_m)^T$ is a $m$-dimensional random vector with $Var(\bm{Z}) = \bm{I}$. We further let $\bm{\Gamma}\bm{\Gamma}^T=\bm{\Sigma}$= $Var(\bm{X})$. The elements $Z_i$'s in $\bm{Z}$ are IID random variables with a pre-specified distribution.  It is worth noting that Model (\ref{eqn:chenM}) is more general than Model (2.4) in \shortciteN{chen2010tests}, where they additionally require that $E(Z_i)=0$ and a uniform bound for the 8th moment for all $Z_i$'s.

We consider  two distributions for $Z_i$: 1) $N(0,1)$; 2) Gamma(4,0.5). \shortciteN{chen2010tests} also considered these two distributions but forcing the mean of Gamma(4,0.5) to be zero in order to meet their requirement of data generation model for their sphericity test.

To evaluate the empirical type I error, we set $\bm{\Gamma} = \bm{I}$  to simulate data under the null  with $\bm{\Sigma}=\bm{I}$.    To evaluate the power, following \shortciteN{chen2010tests}, we consider two forms of alternatives: 1) $\bm{\Sigma} = \sigma_0^2\bm{I} + \sigma_1^2 \bm{A}$, where $\bm{A} = \mathrm{diag}(\bm{I}_{\lfloor\lambda p\rfloor}, \bm{0}_{p-\lfloor\lambda p \rfloor})$,  $\sigma_0^2 = \sigma_1^2 =1$ and  $\lfloor\cdot\rfloor$ is the floor function. In this case $\bm{\Gamma}=\text{diag}\left(\sqrt{2}\bm{I}_{\lfloor\lambda p\rfloor},\bm{I}_{p-\lfloor\lambda p \rfloor})\right)$ and $\bm{Z}$ is a $p \times 1$ random vector with elements that are IID $N(0,1)$ or Gamma(4,0.5) random variables.   2) $\bm{\Sigma} = (1- \rho)\sigma_0^2\bm{ I} + \rho \sigma_1^2\bm{J}$ with  $\sigma_0^2 = 1$ and $\sigma_1^2 =2$. In this case $\bm{\Gamma}=\text{diag}\left(\sqrt{1-\rho}\bm{I},\sqrt{2}\bm{1} \right)$
and  $\bm{Z}$ is a $(p+1) \times 1$ random vector with elements that are IID $N(0,1)$ or Gamma(4,0.5) random variables. We choose the challenging cases in \shortciteN{chen2010tests} by setting $\lambda =0.125$ in the first alternative and $\rho=0.1$ in the second, where the proposed tests of \shortciteN{chen2010tests} had lower power in their simulation study.

We apply the sphericity test (\ref{eqn:sph_var}) to a collection of  simulated datasets with the varying samples sizes and dimensions as considered in \shortciteN{chen2010tests}, and compute p-values with 100 permutations using Algorithm \ref{alg:sph}.  We replicate each simulation setting 2000 times and reject the null hypothesis at the significance level of 0.05. Table \ref{tab:sph_null} shows that the proposed sphericity test produces p-values compatible with the nominal level of 0.05 under the null. Table \ref{tab:sph_pw1} presents the empirical power of our proposed sphericity test under the first alternative, where for a fixed dimension, the empirical power increases quickly with the sample size for both  distributions. For a fixed sample size,  except for the case with the sample size of twenty, the empirical power increases with the dimension.  Table \ref{tab:sph_p2} shows that our proposed sphericity test demonstrates superb empirical power, almost 100$\%$,  under the second alternative. 

\begin{landscape}
	\begin{longtable}{ccccc|cccc}
		\caption{Empirical type I error  (\%) of the sphericity test at the 5\% significance level}  \label{tab:sph_null}\\
		%\hline Sample Size ($n$)& \multicolumn{5}{c|}{$n=100$} &	\multicolumn{6}{c}{$n=4$} \\	
		\hline
		Sample Size ($n$) & 20  & 40  & 60  & 80  & 20 & 40 & 60 & 80  \\
		\hline  
		Dimension ($p$) & \multicolumn{4}{c|}{$N(0,1)$ } &	\multicolumn{4}{c}{Gamma(4, 0.5)}\\
		\hline
		
		38              &   4.9 & 5.2 & 5.1& 5.2& 6.2  & 5.3  & 4.6 & 4.7                               \\
		55             &   4.8& 5.6& 5.0& 4.5 &   4.7 & 5.6 & 4.2  & 4.7                                   \\
		89            &   5.2& 5.1& 4.5& 5.0&     5.0 & 5.2& 4.5 & 4.9                             \\
		159          &    4.8& 5.2 &5.3& 4.2 &    6.0 & 5.3 & 4.5 & 5.1                                  \\
		181           &    5.2& 5.6& 5.0& 4.4&    4.9 & 4.4 & 4.7 & 5.9                            \\
		331          &   4.4& 4.6& 5.0& 4.1&      4.9 &5.5&5.1 & 4.7                                  \\
		343           &   5.3& 3.4 &5.0& 5.4 &     5.8 & 4.2 &5.8 & 4.2                               \\
		642           &    4.8& 4.9& 5.4& 4.8&     5.5& 4.4 & 5.6 & 4.9                                  \\
		
		\hline
		
	\end{longtable}
\end{landscape}

\begin{landscape}
	\begin{longtable}{ccccc|cccc}
		\caption{Empirical power (\%) of the sphericity test of $H_0: \bm{\Sigma}= \sigma^2\bm{I}$ vs $H_1: \bm{\Sigma} = \sigma_0^2\bm{I} + \sigma_1^2\bm{A}$\\ at the 5\% significance level with $\bm{A}=\mathrm{diag}(\bm{I}_{\left[\lambda p\right]},\bm{0}_{p-\left[\lambda p\right]})$, $\sigma_0^2=\sigma_1^2=1$, and $\lambda=0.125$. }  \label{tab:sph_pw1}\\
		%\hline Sample Size ($n$)& \multicolumn{5}{c|}{$n=100$} &	\multicolumn{6}{c}{$n=4$} \\	
		\hline
		Sample Size ($n$) & 20  & 40  & 60  & 80  & 20 & 40 & 60 & 80  \\
		\hline  
		Dimension ($p$) & \multicolumn{4}{c|}{$N(0,1)$ } &	\multicolumn{4}{c}{Gamma(4, 0.5)}\\
		\hline
		
		38              &   32 & 66 & 91& 98&  17  & 48  & 73 & 90                               \\
		55             &    32 & 73& 94& 99 &  16 & 52 & 83  & 95                                   \\
		89             &   36& 83& 98& 100&    17 & 64& 92 & 98                             \\
		159          &   37& 85 &99&100 &    21 &  67& 95 & 100                                  \\
		181          &   37& 86& 99& 100&   20 & 68 & 96 & 100                            \\
		331          &   38& 90& 100& 100&  20 &74&97 & 100                                  \\
		343          &   38& 88 &99& 100 &    21 & 73 &97 & 100                               \\
		642         &    38& 89& 100& 100&   22&73 & 98 & 100                                  \\
		
		\hline
		
	\end{longtable}
\end{landscape}

\begin{landscape}
	\begin{longtable}{ccccc|cccc}
		\caption{Empirical power  (\%) of the sphericity test $H_0:\bm{\Sigma}= \sigma^2\bm{I}$ vs $H_1: \bm{\Sigma} = (1-\rho)\sigma_0^2 \bm{I} + \rho\sigma_1^2 \bm{J}$\\ at the 5\% significance level with $\rho=0.10$, $\sigma_0^2=1$ and  $\sigma_1^2=2$  }  \label{tab:sph_p2}\\
		%\hline Sample Size ($n$)& \multicolumn{5}{c|}{$n=100$} &	\multicolumn{6}{c}{$n=4$} \\	
		\hline
		Sample Size ($n$) & 20  & 40  & 60  & 80  & 20 & 40 & 60 & 80  \\
		\hline  
		Dimension ($p$) & \multicolumn{4}{c|}{$N(0,1)$ } &	\multicolumn{4}{c}{Gamma(4, 0.5)}\\
		\hline
		
		38              &   97 & 100 & 100 & 100&  93  & 100  & 100 & 100                               \\
		55             &    99 & 100& 100 & 100 &  97 & 100 & 100  & 100                                   \\
		89             &   100& 100& 100& 100&    99 & 100& 100 & 100                             \\
		159          &   100& 100 &100&100 &    100 &  100& 100 & 100                                  \\
		181          &   100& 100& 100& 100&    100& 100 & 100 & 100                            \\
		331          &   100& 100& 100& 100&    100 &100&100 & 100                                  \\
		343          &   100& 100 &100& 100 &   100 & 100 &100 & 100                               \\
		642         &    100& 100& 100& 100&    100&100 & 100 & 100                                  \\
		
		\hline
		
	\end{longtable}
\end{landscape}

%In this case we use Algorithm \ref{alg:idt} to compute the p-value.  
\subsubsection{Identity test} \label{identity}

In this section we evaluate the performance of the proposed identity test (\ref{eqn:sph_var}) for covariance or correlation matrices. We design a block-diagonal model for a $p \times 1$ random vector $\bm{X}= (\bm{X}_1^T, \bm{X}_2^T, \bm{X}_3^T, \bm{X}_4^T)^T$, where $\bm{X}_i$ is a $q \times 1$ sub-vector, $i= 1,2,3,4$, such that $Var(\bm{X})=\bm{\Sigma}$, $Var(\bm{X}_i)=\bm{\Sigma}_i$ and $Cov(\bm{X}_i, \bm{X}_j)=\bm{0}$ for $i \neq j$. The covariance matrix in this block-diagonal model takes the form of $\bm{\Sigma} = \text{diag}( \bm{\Sigma}_1, \cdots, \bm{\Sigma}_4)$, where $\bm{\Sigma}_i$ is a $q\times q $ matrix for all $i$'s and $\bm{\Sigma}$ has dimensions of  $p \times p$ with $p = 4q$. We first simulate each $\bm{X}_i=\bm{L}\bm{U}_i$, where $\bm{U}_i$ is a $q \times 1$ random vector and $\bm{L}$ is a lower triangle matrix from the Cholesky decomposition $\bm{\tilde{\Sigma}_q}= \bm{L}\bm{L}^T$, where $\bm{\tilde{\Sigma}_q}$ has a structure of  $(1-\rho)\bm{I}+\rho\bm{J}$. Thus $\bm{\Sigma}_i=\bm{L}\bm{\Lambda}_i\bm{L}^T$, where $\bm{\Lambda}_i$ is a diagonal matrix whose diagonal elements are the variances of the elements of $\bm{U}_i$. Then these four $\bm{X}_i$'s are stacked to obtain the random vector $\bm{X}$.

We consider two configurations for $\bm{X}$. In the first configuration, the four $\bm{U}_i$'s are IID and the elements in all $\bm{U}_i$'s have one of the following four distributions: 1)   $N(0,1)$; 2)  Log-normal(0,1); 3) Student  $t_5$; 4) Gumbel(10, 2) 
%\begin{enumerate}
%\item  $N(0,1)$
%\item Log-normal(0,1)
%\item Student  $t_5$
%\item Gumbel(10, 2) 
%\end{enumerate}
to generate the random vector $\bm{X}$.   In the second configuration,  the four $\bm{U}_i$'s are independently, but not identically, distributed and each $\bm{U}_i$ has a different  distribution. The second configuration is referred as a hybrid configuration.  The reason to choose these four distributions is to investigate the impact of different types of distributions, such as asymmetric distributions, heavy-tail distributions, and extreme value distributions, on the performance of the proposed identity test (\ref{eqn:sph_var}) and its associated  Algorithm \ref{alg:idt}.

We also consider two scenarios for the sample size: 1) a low-dimensional case with $n>p $; 2) a high-dimensional case with $n \ll p$. Using the block-diagonal model, we generate data under the null with $\rho=0$ to evaluate the empirical type I error and under the alternative with $\rho=0.15$ to evaluate the power. Using Algorithm \ref{alg:idt}, we compute p-values based on 100 permutations and reject the null hypothesis at the significance level of 0.05. We replicate each simulation setting 2000 times.   Table \ref{tab:idt} summarizes the results of the empirical type I error and power of  our proposed identity test. It is obvious that the p-values are compatible with the nominal level of  0.05 under the null. The empirical power is close to 100\% when the sample size is bigger than the dimension of the data. When the sample size is much smaller than the dimension, the empirical power increases with the increase of the dimension for this block-diagonal model, though there are some variations across different distributions.

\begin{landscape}
	\begin{longtable}{rccccc|cccccc}
		\caption{Empirical type I error (\%) and power (\%) of the identity test using block-diagonal model\\
			for $H_0: \bm{\Sigma}= \bm{I}$ and $H_1: \bm{\Sigma}=\bm{I}_4 \otimes \bm{\Sigma}_q$, where $\bm{\Sigma}_q = (1-\rho)\bm{I}+\rho\bm{J}$  with $\rho = 0.15$ }  \label{tab:idt}\\
		\hline Sample Size ($n$)& \multicolumn{5}{c|}{$n=100$} &	\multicolumn{6}{c}{$n=4$} \\	
		\hline
		Dimension ($p$) & 24  & 32  & 64  & 76  & 92  &100 & 200 & 300 & 500 & 700 & 1000 \\
		\hline  
			\vspace*{-3mm}
		& \multicolumn{5}{c|}{Type I error } &	\multicolumn{6}{c}{Type I error}\\
	
		$N(0,1)$             & 5.5  &5.0&5.5&5.0&4.8  &  5.3 & 5.5 & 4.5 & 5.0  &  4.1  & 4.9      \\
		Log-normal(0,1)      & 5.0  &4.2&4.4&4.9& 5.4 & 5.8 & 4.9  & 4.4 & 5.2 & 5.1  & 5.0         \\
		Student  $t_5$      &   4.2&5.3&4.4&4.8&5.1  & 4.8 & 4.9  & 4.7 & 5.0  & 5.3 & 5.5         \\
		Gumbel(10,2)            &   5.0&4.5&5.2&5.0&4.6  & 4.9  & 5.2 & 4.6 &5.2   &5.2 & 5.1           \\
		Hybrid             &  4.6 &4.9&4.2&4.4&5.4   &4.9  &  4.5& 5.1 & 5.2 & 5.1 & 4.6           \\
		\hline
			\vspace*{-3mm}
		& \multicolumn{5}{c|}{Power } &	\multicolumn{6}{c}{Power}\\
		
	    $N(0,1)$                  &99&100 &100  &100&100   &  13   & 24  &  31 & 47   & 57  &70                                            \\
		Log-normal(0,1)      &97 &100 &100 &100&100   &  39  & 69 & 82  &  93&  97& 99                                           \\
		Student  $t_5$       &99 &100&100 &100&100  &  18 & 32 & 43  & 58   &  68 & 80                                             \\
		Gumbel(10,2)          & 99&100 &100 &100&100   & 14 & 28 & 40  & 59 & 68   & 79 \                                            \\
		Hybrid               &99 &100 &100 &100&100    & 22  & 43  & 57   &  74 & 84 & 90                                             \\
			\hline
	\end{longtable}
\end{landscape}

\subsection{Two-sample test} \label{two-sample}

In this section we study the empirical type I error and power  of our proposed two-sample test (\ref{eqn:twov}) for the equality of two covariance matrices. We consider two dimensionality configurations: 1) $\min(n_1, n_2) > p$ and 2) $\max(n_1, n_2) \ll p$, where $n_1$ and $n_2$ are the two sample sizes and $p$ is the dimension. We adopt the block-diagonal model as in Section \ref{identity} to simulate data in these two dimensional configurations.   Under the null hypothesis we set $\rho=0.15$ in both samples and  under the alternative we set $\rho_1=0.15$ for sample one and $\rho_2 = 0.30$ for sample two.  We apply  Algorithm \ref{alg:two} with 100 permutations to compute  p-values and reject the null hypothesis at the significance level of 0.05. We replicate each simulation setting 2000 times.

Table \ref{tab:two} shows that the proposed equality test maintains a good control of the empirical type I error at the nominal level of 0.05 under the null hypothesis regardless of the underlying distributions and data dimensions. 

Note that  75\% of the elements in the covariance matrix are zero under the block-diagonal model. This implies that at least 75\% of the elements of $\bm{\Sigma}_1$ and $\bm{\Sigma}_2$ are the same regardless of the value of $\rho$, which makes it rather difficult  to test the equality of two covariance matrices.  In fact, for the same data, the two-sample $Box's$ $M$-$test$, which approximates the two-sample likelihood ratio test based on a multivariate normal distribution,  produces empirical power: 30\%, 30\%, 66\%, 96\% and 100\% for the sample sizes of $n_1$=$n_2$ =100 and the dimensions $p$= 24, 32, 64, 76, 92, respectively. Clearly, the empirical power is not high even  for the $Box's$ $M$-$test$ based on a correct model. The results of our proposed two-sample tests are reported in Table \ref{tab:two}. For the multivariate normal distribution, our proposed method,  on average, outperforms the two-sample $Box's$ $M$-$test$.  The model with log-normal(0,1) gives the lowest power in the first dimensionality configuration where $n_1$=$n_2$=100 ($>p$).   The model with Gumbel(10, 2)  gives the highest power among the four distributions in both configurations. Furthermore, it is interesting to notice that the model with hybrid distributions gives the second highest power in both dimensionality configurations.  We discover that the empirical power of our proposed two-sample test  is sensitive to the signal-to-noise ratio (SNR) of the underlying distribution, defined as the reciprocal of the coefficient of variation. The model with Gumbel(10,2) has the highest empirical power accompanied with the highest theoretical SNR of 1.70, followed by the model with log-normal(0,1) which has the second highest power with a theoretical SNR of 0.35.  The models with $N(0,1)$ and Student $t_5$ have similar low empirical power and both have the theoretical SNR of zero. We will further investigate the impact of SNR in Section \ref{sect:impact}. 

\begin{landscape}
	\begin{longtable}{rccccc|cccccc}
	%	\captionsetup{justification=centering}
		\caption{Empirical type 1 error and power of two-sample test $H_0:\bm{\Sigma}_1 = \bm{\Sigma}_2$ \\
		with $\rho_1 = \rho_2 =0.15$ vs. $\bm{\Sigma}_1 \neq \bm{\Sigma}_2$ with $\rho_1=0.15$ and $\rho_2=0.30$ }  \label{tab:two}\\
		\hline Sample Size ($n$)& \multicolumn{5}{c|}{$n_1=100$, \;$n_2=100$} &	\multicolumn{6}{c}{$n_1=20$,\;$n_2=20$} \\	
		\hline
		Dimension ($p$) & 24  & 32  & 64  & 76  & 92  &100 & 200 & 300 & 500 & 700 & 1000 \\
		\hline  
		& \multicolumn{5}{c|}{Type I error } &	\multicolumn{6}{c}{Type I error}\\
			\vspace*{-3mm}
		$N(0,1)$            & 5.5  & 4.3& 4.2& 4.8& 4.5  &  4.6 & 5.5 & 4.6 & 5.1  &  4.8  & 4.8      \\
		Log-normal(0,1)      & 4.9  & 5.0 & 4.8 & 5.5& 5.3  & 5.8 & 4.9  & 4.4 & 5.2 & 5.1  & 5.0         \\
		Student  $t_5$      & 5.2 & 4.4 & 4.3& 4.3& 5.3  & 5.0 &5.0  & 4.4 & 4.4  & 5.3 & 3.9         \\
		Gumbel(10,2)           &  5.0 &4.7& 5.4& 5.2& 4.7  & 5.0  & 4.8 & 4.8 &5.0   &4.8 & 5.1           \\
		Hybrid            & 4.8  & 4.4& 4.5& 5.4& 5.6   & 4.5  & 5.4   & 5.0 & 4.3  & 5.4& 5.0            \\
		\hline
		& \multicolumn{5}{c|}{Power } &	\multicolumn{6}{c}{Power}\\
			\vspace*{-3mm}
		$N(0,1)$         & 46& 65 & 90  & 92& 95   &  23   & 30  & 30 & 33   &34 &35                                            \\
		Log-normal(0,1)     &9.5 & 13 & 30 & 42& 50   &  24  & 37& 45  & 47&  49 & 49                                          \\
		Student  $t_5$    &29 & 43 & 81 & 85&  90 &  18 & 28 & 27  & 30   &  33 & 31                                              \\
		Gumbel(10,2)          &90 &100 &100 &100&100   & 100 & 100 & 100  & 100 &100   & 100 \                                            \\
		Hybrid          & 37 & 65 & 99  & 99& 100   & 100  & 100  & 100   &  100 & 100 & 100                                              \\
		
		\hline
	\end{longtable}
\end{landscape}

\subsection{Factors affecting the performance of the proposed two-sample test } \label{sect:impact}

In this section, we probe into some factors that may affect the performance of our proposed two-sample test (\ref{eqn:twov}) and its associated Algorithm \ref{alg:two}. Define a random vector $\bm{X}_1=(X_{11}, \ldots, X_{1p})^T$ with $Var(\bm{X}_1)=\bm{\Sigma}_1$ for the population one and $\bm{X}_2=(X_{21}, \ldots, X_{2p})^T$ with $Var(\bm{X}_2)=\bm{\Sigma}_2$ for the population two. 
We recall the assumption required for the proposed two-sample test as 
\begin{equation} \label{assup:two}
X_{1j}\; \text{and}\; X_{2j} \;  \text{have the same  distribution for all} \;j'\text{s}: j= 1, \cdots, p.
\end{equation}
We first investigate the impact of violating assumption (\ref{assup:two}) on the empirical type I error.  Secondly, we use distributions with different SNR's to examine the sensitivity of the proposed two-sample test  to the change of SNR. Finally,  we study the power of the proposed two-sample test under a sparse alternative as in \citeN{cai2013two}, where the two covariance matrices only differ in a very few and fixed number of off-diagonal elements.
In this section  we focus on the scenarios where the sample size  $n$ is smaller than  the dimension $p$. 

\subsubsection{Impact of assumption (\ref{assup:two})}

To investigate the impact of violating assumption (\ref{assup:two})  on our proposed two-sample test,  we adapt  the moving average  models as in  \citeN{li2012two}. To generate $\bm{X}_1$ and $\bm{X}_2$ under the null hypothesis, we use the model
\begin{equation}
   X_{ij} = Z_{ij} + 2Z_{i(j+1)}, \label{eqn:li1}
\end{equation} 
where $j=1, \cdots, p$ and $i=1,2$ for samples one and two, respectively. The two sequences $\{Z_{1k} \}$ and $\{Z_{2k} \}$, $k=1, \cdots, p+1$, consist of IID random variables. Under the alternative, we generate the sample one using Model (\ref{eqn:li1}) and sample two using the following model
\begin{equation}
X_{2j} = Z_{2j} + 2Z_{2(j+1)} + Z_{2(j+2)}, \label{eqn:li2} 
\end{equation}  
where the sequence $\{Z_{2j}\}$ consists of $(p+2)$ IID random variables.  

%We use  Gamma(4, 0.5) and Gamma(0.5, $\sqrt{2}$) as the distributions for the sequence  $\{Z_{ij}\}$ and choose the smallest sample size of twenty as used in \citeN{li2012two}. We replicate the simulation 2000 times.  In each replicate we compute p-value using 100 permutations in Algorithm \ref{alg:two} and reject the null hypothesis at the significance level of 0.05.  

To examine the impact of violating assumption (\ref{assup:two}) under the null hypothesis,  we set $\{Z_{1k}\}$ to be IID Gamma(4, 0.5) random variables and $\{Z_{2k}\}$ to be IID Gamma(0.5, $\sqrt{2}$) random variables in Model (\ref{eqn:li1}).  We choose the smallest sample size of twenty as used in \citeN{li2012two}. We replicate the simulation 2000 times.  In each replicate we compute p-values using 100 permutations in Algorithm \ref{alg:two} and reject the null hypothesis at the significance level of 0.05.  Table \ref{tab:tab3} shows  that the violation of assumption (\ref{assup:two}) can result in incorrect empirical type I errors in our proposed two-sample test.  On the other hand, when the two samples are generated using the same distribution,  the proposed two-sample test produces correct empirical type I errors compatible with the nominal level of  0.05.   

\subsubsection{Impact of SNR}

To investigate the impact of SNR on the power of our proposed two-sample test, we generate $\bm{X}_1$ using Model (\ref{eqn:li1}) in the sample one and $\bm{X}_2$ using  Model (\ref{eqn:li2}) in sample two. The sequences $\{Z_{1k}\}$ and $\{Z_{2k}\}$  have the same distribution of  either Gamma(4, 0.5) or Gamma(0.5, $\sqrt{2}$).
The two Gamma distributions have the same variance but different means, which imply that they have different SNR's. Table \ref{tab:tab3} shows that the model with Gamma(4, 0.5) has  much higher empirical power than that of the model with Gamma(0.5, $\sqrt{2}$).  This is in line with our observation in Section \ref{two-sample} that higher SNR's result in higher power in our proposed two-sample test, since the model with Gamma(4, 0.5)  has a SNR that is twice  bigger than that of the model with Gamma(0.5, $\sqrt{2}$). To investigate the impact of SNR further,  we consider an additional simulation study with two samples generated using the distributions  of  $N(5,1)$ and  $N(5,5^2)$, respectively, and the former has a SNR that is  five times bigger than the latter.   We use $n_1=n_2=20$  for the sample sizes and $p=50$ for the dimension.   Simulation results in Table \ref{tab:tab3} show that the empirical power of the model with $N(5,1)$ is about 100\%, whereas the power reduces significantly to 27\% using the sample generated by the model with $N(5, 5^2)$.  
 
 \subsubsection{Impact of sparseness}

To investigate the impact of sparse signals, we adopt a model from the supplemental simulation study of \cite{cai2013two}.  Under the null hypothesis, we set $\bm{\Sigma}_1=\bm{\Sigma}_2=\bm{\Sigma}_{null}$, where
\begin{equation} \label{eqn:null}
\bm{\Sigma}_{null}= (\bm{\Sigma}^* + \delta_0 \bm{I})/(1+\delta_0) + \delta_1 \bm{I}.
\end{equation}
Under the alternative, we set $\bm{\Sigma}_1=\bm{\Sigma}_{null}$ and $\bm{\Sigma}_2=\bm{\Sigma}_{alt}$, where

\begin{equation} \label{eqn:salt}
\bm{\Sigma}_{alt}= (\bm{\Sigma}^* + \delta_0 \bm{I})/(1+\delta_0)+ \bm{\Delta} + \delta_1 \bm{I}.
\end{equation}
In models (\ref{eqn:null}) and (\ref{eqn:salt}), $\bm{\Sigma^*}=(\sigma_{ij}^2)$ is a symmetric matrix, in which $\sigma_{ii} =1$ and $\sigma_{ij} =0.5*\text{Bernoulli}(1,0.05)$ for $i <j$. $\delta_0$ = $|\lambda_{\text{min}}(\bm{\Sigma}^*)|$ + 0.05 and $\delta_1=|\text{min}\{\lambda_{\text{min}}((\bm{\Sigma}^* + \delta_0 \bm{I})/(1+\delta_0)),\lambda_{\text{min}}((\bm{\Sigma}^* + \delta_0 \bm{I})/(1+\delta_0) +\bm{\Delta})\}|+0.05$, where $\lambda_{\text{min}}(\cdot)$ returns the smallest eigenvalue of the matrix which it applies to.  $\bm{\Delta}$ is a square matrix containing thirty-two entries of 0.9 and all the other entries are zero. The locations of sixteen nonzero entries are randomly selected in the lower triangle portion and the other sixteen nonzero entries are located in the upper triangle portion accordingly to make $\bm{\Delta}$ symmetric.  Regardless of the dimension of the covariance matrix,  the number of nonzero entries in  $\bm{\Delta}$  remains thirty-two to keep  the signals  sparse. Note that the matrix $\bm{\Delta}$ changes from one replicate to another in this simulation study.

Through the Cholesky decomposition we obtain  $\bm{\Sigma}_{null}=\bm{\Gamma}_1\bm{\Gamma}_1^T$  and $\bm{\Sigma}_{alt}=\bm{\Gamma}_2\bm{\Gamma}_2^T$.  Under the null 
we generate $\bm{X}_1=\bm{\Gamma}_1\bm{U}$ and  $\bm{X}_2=\bm{\Gamma}_1\bm{U}$ to evaluate the empirical type I error, where $\bm{U}$ is a random vector whose elements are IID random variables. Under the alternative we generate $\bm{X}_1=\bm{\Gamma}_1\bm{U}$ and  $\bm{X}_2=\bm{\Gamma}_2\bm{U}$ to evaluate the power of our proposed two-sample test.

Following \citeN{cai2013two}, we set the sample size to thirty for each sample. We use Algorithm \ref{alg:two} to compute p-values with 100 permutations.  We replicate each simulation setting 2000 times. We also study several distributions for the elements in $\bm{U}$. Due to the sampling variations in small samples, the sample SNR's may differ from their corresponding theoretic SNR's. We therefore consider the sample signal-to-noise (SSNR) computed using the sample mean and the sample standard deviation. Table \ref{tab:title} presents the influence of sparse signals and SSNR's on the performance of our proposed two-sample test and its associated Algorithm \ref{alg:two}.  The results show that, under the null hypothesis, our approach produces p-values compatible with the nominal level of 0.05, regardless of the dimension of data and the level of SSNR's. In terms of the empirical power,  our method is not sensitive to different distributions but to the level of SSNR's: the higher SSNR is, the higher the empirical power is.  Moreover, Table  \ref{tab:title}  shows that distributions with similar SSNR's tend to have similar power. 

\begin{landscape}
	\begin{longtable}{rc|cccccc}
		\caption{Empirical type I error (\%) and power  (\%)of the proposed two-sample test}  \label{tab:tab3}\\
		\hline
		\multicolumn{2}{c|}{Dimension $p$} & 50  & 100  & 200  & 300  &600 & 800  \\
		\hline  
		Sample 1 & Sample 2 	& \multicolumn{6}{c}{Type I error } \\
		\hline     
		Gamma(4, 0.5)  & Gamma(0.5, $\sqrt{2}$)     &   100  & 100 & 100  & 100   & 100   & 100       \\	
		Gamma(0.5, $\sqrt{2}$)  & Gamma(0.5, $\sqrt{2}$)     &   5.8& 5.7& 4.3 &5.6   & 5.0  & 5.0      \\
		
		Gamma(4, 0.5)  & Gamma(4, 0.5)     & 5.3   &  5.7 & 6.0  & 5.2   & 5.3  & 5.5       \\
		\hline
		Sample 1 &  Sample 2&	\multicolumn{6}{c}{Power}\\
		\hline
		Gamma(4, 0.5) & Gamma(0.5, $\sqrt{2}$)   &100   & 100  &100  &100    &100   &100                                            \\
		Gamma(0.5, $\sqrt{2}$) & Gamma(0.5, $\sqrt{2}$)   &  13  & 20  &37  & 51    & 87  & 97                                           \\
		
		Gamma(4, 0.5) & Gamma(4, 0.5)   & 99    & 100   & 100   & 100     & 100   & 100                                           \\
		
		\hline
		
		\hline
	\end{longtable}
\end{landscape}

\begin{landscape}	
\begin{longtable}{rcccccc|cccccc }
	\caption{Impact of sparse signal and sample signal-to-noise ratio (SSNR) on the proposed two-sample test} \label{tab:title} \\
    
	\hline
  Dimension ($p$)	&   & 50 & 100 & 200  & 400 & 800  &  & 50 & 100 & 200  & 400 & 800\\
	\hline  
	&  SSNR & \multicolumn{5}{c|}{Type I  error (\%)}  &  SSNR &\multicolumn{5}{c}{power (\%)} \\
	
	$N(0,1)$        & 0.0 (0.0)\textsuperscript{*}  & 5.5  & 5.3 & 4.4  & 5.8  & 5.0  &   0.0 (0.0)   &33   & 17   & 10  & 8  & 7  \\
	$N(2,1)$        & 2.4 (0.1)  & 5.5  & 4.7 & 4.7  & 5.0   & 5.3    & 2.5 (0.1)     & 48 &  27  & 17   & 13   & 9\\
	
	$N(4,1)$        & 4.9 (0.1) & 5.9  & 4.7 & 4.7  & 5.0   &  5.3 & 5.0 (0.1) & 99  &  96  & 89   & 80   & 76 \\
	 
	Gamma(5,1)         & 2.7 (0.1) & 4.6 & 5.3  & 5.3 & 5.0  & 5.3   & 2.8 (0.1) & 55 & 34  &  21   & 15 & 11 \\
	
	Gamma(10,1)    & 3.9 (0.1) & 4.8  & 5.3 &5.4  & 5.6 & 5.2  & 4.0 (0.1) & 89  & 74  &  57   &  42& 36 \\
	
  Poisson(5)   & 2.7 (0.1)  & 5.1  & 5.3  & 4.9 &5.0 & 5.3 & 2.8 (0.1) & 53  &  35 &  22  &14 & 10 \\ 
  Poisson(10)   & 3.8 (0.1)  & 4.7  & 5.1  & 5.2  & 5.4 & 4.3  & 3.9 (0.1) & 88   &  76 &  57   &  42 & 34  \\ 
	Log-normal(0, 0.4) & 3.0 (0.1) & 4.9  & 5.0  &  4.3   & 5.5 & 4.8 & 3.0 (0.1) & 58  & 38  & 25    & 17& 13\\
	Log-normal(0, 0.3) & 4.0 (0.1) & 5.1  & 5.4  & 4.1    &5.8 & 4.9 & 4.1 (0.1) & 91  & 77  & 61   & 47 & 39\\	
  	
	\hline
	\multicolumn{6}{l}{\textsuperscript{*}\footnotesize{Standard deviation of SSNR}}\\
\end{longtable} 

\end{landscape}

%\section{Real Examples}

\section{Discussion} \label{sec:discussion}

In this study we introduce a generalized cosine measure between two symmetric matrices and, based on this geometry, we propose new test statistics for  one-sample and two-sample tests of covariance and correlation matrices.  These test statistics and their companioned algorithms for computing p-values can be applied to a variety of hypothesis tests of covariance and correlation matrices. In the one-sample setting, we implement the test statistics and algorithms for the hypothesis tests of sphericity, identity and compound symmetry. We also propose new test statistics and  algorithms to test the equality of two or multiple covariance or correlation matrices. We demonstrate the effectiveness of these tests and algorithms through several real datasets and extensive simulation studies, where data are generated using a variety of  models and distributions adopted in the previous literature.  

In order for the proposed algorithms to work properly, the number of permutations is important and usually can be determined by the pre-specified significance level.  For instance,  we recommend  at least one hundred permutations for the significance level of 0.05, for this number can provide the required precision to two decimal places.  Moreover, the extensive simulations show that one hundred permutations appear to be sufficient to control the empirical type I error and provide good power for a variety of models and distributions at the significance level of 0.05. If a greater significance level is desirable, for example 0.001 in a multiple testing problem, then at least one thousand permutations is recommended.  Furthermore, the dimension and the sample size of data also impact on the number of recommended permutations. For algorithms associated with the proposed one-sample tests, the dimension plays a more prominent role than the sample size, since the number of permutations is often determined by the number of columns in the data matrix, i.e., the dimension of data. We show in Table \ref{tab:idt} that the proposed algorithm can provide a correct empirical type I error for data with sample size as small as four in identity test, provided that the dimension of  data is sufficiently large.  In the $K$-sample test the sample size may put a restriction on the number of possible permutations, since the algorithms for the $K$-sample tests randomly shuffle rows of the stacked data matrix. For example, in the two-sample test, the minimum recommended sample size for the combined sample is ten with the smaller sample containing at least three subjects, for $10\choose 3$ is 120 and exceeds the  number of permutations recommended for the significance level of 0.05. Generally for a $K$-sample test, we recommend that ${n\choose n_{(1)}} \ge k_0$, where $n_{(1)}$ is the smallest sample size of the $K$ samples and $k_0$ is the required number of permutations for a pre-specified significance level.

Our proposed test statistics are general and flexible so that they can be easily extended to handle other complex situations. As an example, we outline one possible extension as follows. It is often of interest to test the independence of several normal random vectors as in \citeN{fujikoshi2011multivariate} or more generally uncorrelation of multiple random vectors.  Suppose a random vector $\bm{X}$ can be partitioned into $k$ subvectors of $\bm{X}_1, \ldots, \bm{X}_k$ with the length of $p_1, \ldots, p_k$, respectively.  Below we let $k=2$ for an illustrative purpose.  Let the covariance matrix of $\bm{X}$ be
\begin{equation}
\Sigma=
\begin{bmatrix}
\Sigma_{\bm{X}_1\bm{X}_1}   & \Sigma_{\bm{X}_1\bm{X}_2}  \\
\Sigma_{\bm{X}_2\bm{X}_1}  &  \Sigma_{\bm{X}_2\bm{X}_2} \\  
\end{bmatrix},
\end{equation}
where $\Sigma_{\bm{X}_i\bm{X}_j}$ is the covariance matrix of $\bm{X}_i$ and $\bm{X}_j$.  The null hypothesis of uncorrelation between $\bm{X}_1$ and $\bm{X}_2$ is
\begin{eqnarray}
H_0:  \Sigma=
\begin{bmatrix}
\Sigma_{\bm{X}_1\bm{X}_1}   & \bm{0}  \\
\bm{0} &  \Sigma_{\bm{X}_2\bm{X}_2} \\  
\end{bmatrix}.
\end{eqnarray} 
To test this null hypothesis, we specify $\bm{\Sigma}_0$ in the one-sample test statistic  (\ref{eqn:one_sample}) as 
\begin{equation*}
\bm{\Sigma}_0 = \mathrm{diag}(\bm{J}_1, \bm{J}_2),
\end{equation*}
where $\bm{J}_i$ is a $p_i \times p_i $ matrix of ones for $i=1,2$, respectively.  Using the mapping $\vech(\cdot)$, the Euclidean inner product and norm,  it can be shown that test statistic (\ref{eqn:one_sample}) in this case  can be expressed as 

\begin{equation} \label{eqn:svb}
1- \frac{\vech(\bm{S})^T\vech(\bm{\Sigma}_0)}{\sqrt{0.5\sum_{i=1}^{2}p_i(p_i+1)}\|\vech(\bm{S})\|},
\end{equation}
where $\bm{S}$ is the sample covariance matrix. Below we outline an associated algorithm to compute the p-value of (\ref{eqn:svb}) under the null hypothesis of uncorrelation.  Analogous to the identity test, we consider permutations of the elements in each column of the data matrix. In this case the data matrix has two ``group-column"'s: the first ``group-column'' consists of the  first $p_1$ columns in the data matrix  and the second ``group-column" includes the remaining $p_2$ columns.  We then permute the elements in each of the ``group-column"'s to compute the p-value under the null, for this permutation procedure retains the sample covariance matrices for each of the two sub-vectors, while disrupting the covariance between the two.    In future study we  plan to extend the proposed tests and permutation algorithms to handle other covariance structures, for example, those frequently adopted in repeated measure data analysis such as AR(1) and Toeplitz. 

\section{Acknowledgement}
    We would like to thank the computing facility SHARCNET (Ontario, Canada) for a partial computation support.

\section*{Appendix}
\subsection*{Appendix I Derivation of sphericity test statistic (\ref{eqn:sph_var}) and identity test statistic (\ref{eqn:idt_R})} 
Let $\bm{S}$ be a $p \times p$ covariance matrix and $\bm{I}$ a $p \times p$ identity matrix. 
For the sphericity test, $\bm{\Sigma}_0$ in the one-sample test statistic (\ref{eqn:one_sample}) is $\sigma^2\bm{I}$. We adopt the mapping $\vech(\cdot)$, the Euclidean inner product and norm and apply Equation (\ref{eqn:def}) to compute $cos(\bm{S}, \sigma^2\bm{I})$  as follows:
\begin{align}
cos(\bm{S},  \sigma^2\bm{ I}) &= \frac{\vech(\bm{S})^T\vech(\sigma^2\bm{I})}{\|\vech(\bm{S})\|\|\vech(\sigma^2\bm{I})\|} \notag \\
&=\frac{\vech(\bm{S})^T\vech(\bm{I})}{\|\vech(\bm{S})\|\|\vech(\bm{I})\|}  \notag \\
&= \frac{\sum_{i=1}^{p}s_{ii}^2}{\sqrt{p}\|\vech(\bm{S})\|} \notag \\
&=\frac{\tr(\bm{S})}{\sqrt{p}\|\vech(\bm{S})\|}, \label{eqn:trace}
\end{align}
where $s_{ii}^2$, $i=1, \ldots, p$,  are the diagonal elements of $\bm{S}$. Then the sphericity test statistic (\ref{eqn:sph_var}) follows from $1- cos(\bm{S}, \sigma^2\bm{I})$.

For the identity test of a correlation matrix $\bm{S}$, the $\bm{\Sigma}_0$ in the one-sample test statistic (\ref{eqn:one_sample}) is $\bm{I}$.  We adopt the mapping $\vech(\cdot)$, the Euclidean inner product and norm and apply Equation (\ref{eqn:def}) to compute $cos(\bm{S}, \bm{I})$, which leads to equation (\ref{eqn:trace}). Since $\bm{S}$ is a sample correlation matrix and $\tr(\bm{S})$=$p$, the $cos(\bm{S}, \bm{I})$ can be simplified into
\begin{equation}
cos(\bm{S}, \bm{I}) = \frac{\sqrt{p}}{\|\vech(\bm{S})\|},
\end{equation}
then the identity test statistic (\ref{eqn:idt_R}) follows from $1-cos(\bm{S}, \bm{I})$.

\subsection*{Appendix II Derivation of the compound symmetry test statistic (\ref{eqn:com})} 
Let $\bm{S}$ be a $p \times p$ sample covariance or correlation matrix and $\bm{J}$ a $p \times p$ matrix of one's. For the compound symmetry test, $\bm{\Sigma}_0$ in the one-sample test statistic (\ref{eqn:one_sample}) is $\sigma^2[(1-\rho)\bm{I}+\rho\bm{J}]$ for the covariance matrix and  $[(1-\rho)\bm{I}+\rho\bm{J}]$ for the correlation matrix, where $\rho>0$ is the intra-class coefficient coefficient.  We first derive the test statistic for the covariance matrix and show that it does not depend on the unknown parameters of $\sigma^2$ and $\rho$.

We adopt the modified half-vectorization $\vech^*(\cdot)$,  the Euclidean inner product and norm to compute $cos(\bm{S}, \sigma^2[(1-\rho)\bm{I}+\rho\bm{J}] )$ according to Equation (\ref{eqn:def}) as follows:

\begin{align}
cos(\bm{S},  \sigma^2[(1-\rho)\bm{I}+\rho\bm{J}] ) &= \frac{\vech^*(\bm{S})^T\vech^*( \sigma^2[(1-\rho)\bm{I}+\rho\bm{J}] }{\|\vech^*(\bm{S})\|\|\vech^*(\sigma^2[(1-\rho)\bm{I}+\rho\bm{J}])\|} \notag \\
&=\frac{\vech^*(\bm{S})^T\vech^*(\bm{J})}{\|\vech^*(\bm{S})\|\|\vech^*(\bm{J})\|}  \notag \\
&=\frac{\vech^*(\bm{S})^T\vech^*(\bm{J})}{\sqrt{0.5p(p-1)}\|\vech^*(\bm{S})\|}. \label{eqn:comTest}
\end{align}
which  does not involve the unknown parameters of $\sigma_2$ and $\rho$.    The test statistic (\ref{eqn:com}) for the covariance matrix $\bm{S}$ follows from  $1-cos(\bm{S},  \sigma^2[(1-\rho)\bm{I}+\rho\bm{J}])$.

Suppose $\bm{S}$ is a sample correlation matrix,  we adopt the modified half-vectorization $\vech^*(\cdot)$,  the Euclidean inner product and norm to compute $cos(\bm{S}, [(1-\rho)\bm{I}+\rho\bm{J}] )$ according to Equation (\ref{eqn:def}).  Following the above calculation for the sample covariance matrix, we can show that $cos(\bm{S}, [(1-\rho)\bm{I}+\rho\bm{J})])$ yields the same equation (\ref{eqn:comTest}) and the test statistic (\ref{eqn:com})  follows from $1-cos(\bm{S},  [(1-\rho)\bm{I}+\rho\bm{J}])$.

\bibliography{junk}

\bibliographystyle{mychicago}

\end{document}